# Exploring k-Colorability
## by Kia Kai Li
kia.kai.li@gmail.com

**Introduction**

The P vs NP problem is now well-known worldwide, even beyond academia, due to the publicity generated by the Clay Mathematics Institute, which bestowed upon it one of seven Millennium Prizes on a correct proof of the solution. The concept of NP-completeness was first formulated by Stephen Cook, in application to the boolean satisfiability problem [Coo71]. Later, Richard Karp took 20 additional diverse natural problems, each known to be hard to solve, and showed them to be related by NP-completeness [Kar72]. Since then, thousands of problems have shown to belong in this theoretical class in computer science.

One of the more popular NP-complete problems is graph k-colorability, a problem that has a relatively simple definition, and yet has defied solution since it was formulated. Following the common definition, a graph is k-colorable if each vertex has one color different from those of all its adjacent vertices, those connected directly to said vertex with an edge, such that when the whole graph is colored, only k or less colors have been used. As this paper will show, this seemingly simple problem has resisted solution due to its unexplored intrinsic structure and properties. This is true even with only 3 colors.

**Background**

A graph G = [V,E] with V vertices or nodes connected by E edges is said to be k-colorable if for each node $v_i$ with color $c_i <= k$, each node $v_j$ adjacent to $v_i$ has color $c_j <= k$ such that $c_j$ and $c_i$ are different colors. The chromatic number *chi(G)* is the minimum k for which G is colorable. For those unfamiliar with graphs in computer science, the nodes are simply zero-dimensional point objects, while the edges are one-dimensional line segments connecting the nodes. Only simple undirected graphs are considered here, such that the edges are nonredundant and undirected non-loops. The colors, being an extrinsic property of the graph, can be conveniently expressed as natural numbers, which would facilitate their assignment in a computer algorithm. The *degree of freedom* of a node during the run of the algorithm is the number of colors still available for that node based on the colors of the adjacent nodes.

The efficiency of an algorithm is best characterized by the "big O" notation. First, the input size of the problem is determined. The dimension of the input graph is O(V) in the number of vertices and O(E) = $O(V^2)$ in the number of edges. If one can show that a constant number of computational steps is needed to color each node, then the computational costs of the algorithm is said to grow in proportion to the number of nodes. In complexity theory terms, the runtime of the algorithm grows linearly on the input size V, and the problem is said to belong to class P.

Of course, many problems that resist efficient solutions may contain in their solution space an ordered structure that has yet been discovered. If the local structure exhibits a direct relationship to the global structure, then a greedy algorithm would likely lead to a solution. But the existence of local structure also implies that the number of computational steps would vary as determined by the state of the algorithm within the solution space. In complexity theory, the number of computational steps needs only to be polynomially bounded to the input size for the problem to be considered to reside in the tractable class P. As shown in a later section, determining whether a graph can be colored with only 2 colors can be done in polynomial time, or *PTIME*. Therefore, 2-colorability is in class P.

The class NP has only one property in common with the class P, which is that given the solution, it can be efficiently verified in polynomial time. In fact, this is one definition of the class NP. Thus, the class NP is said to contain the class P. Finding the solution to NP-complete problems from scratch requires brute-force methods that take exponential time, or *EXPTIME*.  No one has yet been able to find efficient solutions for them, leading many, including 61 out of 100 researchers in one recent poll [Gas02], to believe the two classes to be different.  In particular, it is not known how to k-color a graph exactly, with k >= 3, in polynomial time.  Hence, 3-colorability is in class NP, but being intractable, it resides outside of class P.  In fact, it has been proven that 3-colorability on planar graphs with maximum degree of 4 is NP-complete [Sto73].  An NP-complete problem is among the hardest in the NP class, and can be reduced in polynomial time from and to other NPC problems.

**Exploring Graph Theory**

It is useful to understand the context of a problem before trying to find its solution.  In trying to understand graphs as an abstract representation used primarily in computer science, one would find the study of graph theory quite different from commonly known mathematical disciplines.  Proofs in graph theory are more logical than formulaic as in, e.g. calculus or differential equations.

There are quite a few topics in graph theory, including paths, trees, and many other specific graph classes and types.  Although they has been studied to better understand the coloring problem, most of the respective work is omitted from this paper to keep the focus on the ultimate goal of a generic exact graph coloring algorithm.  A graph coloring handbook would be an useful source [JT95].  There is also not enough space here to properly introduce graph theory, so here suggests an up-to-date source [Wal00] as well as a historical account spanning the last 200 years [BLW76].

It should be noted that not all graph theory topics are relevant to graph coloring.  For example, subdivision may be important in determining graph planarity, but may not be particularly helpful to graph colorability since the decision problem absolutely changes on relaxation of even one edgewise constraint.  Similarly, it is unclear how graph contractions affect graph coloring.

One topic worth mentioning in relation to graph coloring is the degree of a vertex, represented by $d(v_i)$.  The degree of a vertex is the number of edges incident to that vertex.  The maximum degree of a graph G is represented by *Delta(G)*.  Assuming all $d(v_i)$ adjacent nodes of a vertex $v_i$ have different colors, taking up $d(v_i)$ colors, $v_i$ must take the $d(v_i)$+1st color.  Applying this to the node with maximum degree Delta in a graph G, *chi(G) <= Delta(G) + 1*.

**2-Colorability**

To determine if a graph can be colored with 2 colors, one only needs to determine if the graph contains any odd loops.  One implementation of this property involves the use of DFI, or depth-first index.  DFI derives from depth-first search, in which each deeper level in the tree contains an index incremented by one from the previous level.  Starting from any node, run DFS and assign DFI accordingly.  An odd loop will reveal itself when two nodes at the same level are found to be adjacent to each other.  Since DFS travels along the edges, it is $O(E) = O(V^2)$, with runtime polynomial in the size V of the input graph.  Therefore, 2-colorability is in P.  A proof connecting the presence of an odd loop to 2-colorability follows.

Proof.  In trying to 2-color a graph, any two adjacent nodes must take up both colors.  One can then see that every other node must be the same color in a 2-color graph.  Let's call the colors "odd" and "even".  Just like in the number line, where odd and even numbers alternate, the two colors, "odd" and "even", can alternate without conflict in a graph representing the number line.  This remains valid even when there are multiple number lines intersecting at only one point and at no other nodes.  But when there are two points of intersection, a loop is formed.  It is useful to compare the loop object to the face of a clock.  A typical clock face has 12 numbers on it.  Tracing those numbers produce an even loop with 12 nodes and 12 edges.  The loop can be 2-colored because the odd and even alternate on the original clock face.  To produce an odd loop, one takes away one node and an edge adjacent to that node, and then reconnect the loop.  In the clock example, taking away the number 12 would make 1 and 11 adjacent to each other.  The algorithm thus has failed by allowing two "odd" colors to be adjacent to each other.  Finally, it is easy to see that paths and loops of any length completely build up to any well-defined simple graphs.  QED

## 3-Colorability

It is truly mind-boggling how adding a third color can push the complexity of the problem into a whole different class. It is unclear how the additional complexity is distributed, but it is relatively simple to explain its source. Remember two colors suffice to take both ends of an edge. But by allowing a third color, the color of an adjacent node is no longer enough to determine the color of the current node, since two possible colors remain. One could try using the smallest color available. Coloring the graph in this way yields the Grundy number [Gru39], which is at least as large as the chromatic number, thus, leaving the coloring problem still unsolved.

Going back to the easier problem of 2-colorability, it is determined that the odd loop is the inhibiting component of a non-2-colorable graph. So if an odd loop can be 3-colored, what is the inhibiting component of a non-3-colorable graph? Imagine this line of question for k going up to infinity. One would naturally want to try to distinguish a basic rule or pattern for coloring in general. One such rule is that a single edge can make the difference between colorability. The edge would connect two nodes of the same color, with no degrees of freedom among them. Based on this idea, a different picture of 3-colorability can be constructed. Assuming one node has the third color in an odd loop, call that node the "hole". The idea is to rotate the holes in the odd loops such that they are not adjacent to each other. The problem then becomes isolating the holes and 2-coloring the rest of the graph.

But the relationship between the holes and the rest of the graph is not certain and cannot yet be proven. This again brings up the issue of local structure. With respect to 3-colorability, how do holes relate to adjacent nodes and beyond? Can holes be adjacent to each other without causing conflict in coloring? Assuming the answer to the latter is affirmative, it becomes clear that the previous definition of the third color filling the "hole" is too narrow. Looking at the bigger picture, the 3 colors only need to be arranged optimally at the local level in order for the graph to be 3-colorable. i.e. The third color can occupy the "hole" in one section of the graph, while the first or second color could assume that role in a different section. To put it more succinctly, all 3 colors must balance each other in all of the odd loops for the graph to be 3-colorable.

It now seems likely that such local structure exists. It can conceivably be applied for any k, the result dependent on the k being tested for and the graph instance being tested on. Indeed, the colorability problem retains its nature as k increases. Any work toward discerning such rules or structure would surely bring progress toward the solution of the graph k-colorability problem.

## Beyond k-colorability

The k-colorability problem has several important real-world applications, including register allocation, scheduling, frequency assignment, and many other problems in which an enumerable resource is distributed based on given pairwise constraints. The problem also underlies various popular games, including Sudoku and Minesweeper.

Given the ubiquity of NP-completeness, it is not surprising to find the nature of graph colorability present in other problems as well. The most closely related would be the clique problem and the independent set problem. The clique problem tries to find the maximum clique of size *omega(G)*. A clique is a set of nodes that are all connected to each other. The independent set problem tries to find the maximum independent set of size *alpha(G)*. An independent set is a set of nodes in which none of the nodes are connected to each other. One interesting connection among these problems is as follows: *omega(G) <= chi(G) <= alpha(G)*.

Another NP-complete problem is the vertex cover problem, which asks for the smallest set of nodes of size *MVC(G)* such that all edges of G have at least one endpoint in the set. Interestingly, while the vertex cover problem is very similar to the other NPC problems mentioned previously, the attribute it embodies, *MVC(G),* can be smaller than *omega(G)* or bigger than *alpha(G)* depending on the graph instance. Such relationships are good evidence of much unexplored structure that remains even in simple generic graphs.

Finally, reducibility between NP-complete problems means that any structural rules in graph colorability can be ported into solutions for all other NP-complete problems, regardless of how disparate and distinct the qualities those problems may seem to possess. Thus, future work can be directed toward discerning structure in the solution space of all remaining NP-complete problems. For example, the boolean satisfiability problem can be reduced by assigning colors to the values of TRUE, FALSE, and a dummy boolean value, to all variables and their complements, and to the specific clauses that comprise each problem instance. The subset sum problem has been reduced by breaking down the relationship among the digits of each number in the input set.

**Conclusion**

Having the property of NP-completeness, k-colorability still holds much interest in theoretical computer science. Many in the industry can also appreciate the connection of the intractability of this problem class to real-world applications. Yet instead of distinguishing between abstract theory and real-world problem solving, the two should be considered as flip sides of the same coin, complementing each other, if we are to expect progress. From one perspective, the only difference between theory and its application is the language used in their descriptions. But how can one determine the limits of a language without fully understanding its structure? The P vs NP problem represents such a limit of our times. As mankind has done to overcome all other barriers, computer scientists and mathematicians alike must continue to establish all foundations within the broad expanse of their fields, not only toward the goal of discovering central unifying principles of our universe, but for the ultimate completeness in human knowledge.